\documentclass[10pt, twocolumn, floatfix, superscriptaddress, aps, prl]{revtex4-2}

\usepackage{amsmath, amssymb, graphicx, hyperref, xcolor, colortbl, braket}
\usepackage{cleveref}
\usepackage{comment}
\usepackage{mathtools}

\DeclareMathOperator{\tr}{tr}
\DeclareMathOperator{\diag}{diag}

\begin{document}

\title{Stable infinite-temperature eigenstates in SU(2)-symmetric nonintegrable models}

\author{Christopher J. Turner}
\affiliation{Department of Physics and Astronomy, University College London, Gower Street, London, WC1E 6BT, UK}

\author{Marcin Szyniszewski}
\affiliation{Department of Physics and Astronomy, University College London, Gower Street, London, WC1E 6BT, UK}

\author{Bhaskar Mukherjee}
\affiliation{Department of Physics and Astronomy, University College London, Gower Street, London, WC1E 6BT, UK}

\author{Ronald Melendrez}
\affiliation{Department of Physics, Florida State University, Tallahassee, Florida 32306, USA}
\affiliation{National High Magnetic Field Laboratory, Tallahassee, Florida 32310, USA}

\author{Hitesh J. Changlani}
\affiliation{Department of Physics, Florida State University, Tallahassee, Florida 32306, USA}
\affiliation{National High Magnetic Field Laboratory, Tallahassee, Florida 32310, USA}

\author{Arijeet Pal}
\affiliation{Department of Physics and Astronomy, University College London, Gower Street, London, WC1E 6BT, UK}

\begin{abstract}
Nonintegrable many-body quantum systems typically thermalize at long times through the mechanism of quantum chaos. However, some exceptional systems, such as those harboring quantum scars, break thermalization, serving as testbeds for foundational problems of quantum statistical physics. Here, we investigate a class of nonintegrable bond-staggered models that is endowed with a large number of zero-energy eigenstates and possesses a non-Abelian internal symmetry. We use character theory to give a lower bound on the zero-energy degeneracy, which matches exact diagonalization results, and is found to grow exponentially with the system size. We also show that few-magnon zero-energy states have an exact analytical description, allowing us to build a basis of low-entangled fixed-separation states, which is stable to most perturbations found in experiments. This remarkable dynamical stability of special states elucidates our understanding of nonequilibrium processes in non-Abelian chaotic quantum models.
\end{abstract}

{\maketitle}

\textit{Introduction.}---
The concept of nonintegrability is closely associated with the thermalization of all eigenstates of a many-body quantum system, alongside their complex chaotic behavior~\cite{Deutsch1991, Srednicki1994, DAlessio2016, Borgonovi2016}. At long times, the average properties of such systems can be described using the microcanonical ensemble of statistical mechanics, effectively erasing all information about initial conditions to local observables~\cite{Deutsch1991, Jensen1985, Rigol2012}. Despite this statistical description, nonintegrable models remain analytically intractable~\cite{Rigol2008, Kim2014, Beugeling2014}: unlike integrable systems, they lack an extensive set of local integrals of motion~\cite{Shiraishi_2019}, leaving numerical treatments our primary source of understanding.

However, certain nonintegrable models have recently been found to defy such a statistical description.
Examples include systems with many-body quantum scars~\cite{Moudgalya2018, Moudgalya2018scars, Turner2018nat, Turner2018, Lin2019, Schecter2019, Choi2019, Iadecola2020, Lee2020}, which are atypical states within the spectrum which violate the eigenstate thermalization hypothesis and sometimes can be constructed exactly.
Often these are characterized by low entanglement, but sometimes occur with volume-law entanglement~\cite{Langlett2022, Wildeboer2022, Ivanov2024Mar, Chiba2024}.
These out-of-equilibrium states are gaining interest due to their potential applications for quantum computing architectures~\cite{Morvan2022Dec}.

Interestingly, the existence of degenerate manifolds can also signify potential analytical tractability, with applications spanning low energy physics~\cite{Changlani_PRL2018}, the order by disorder mechanism in frustrated magnetism~\cite{orderbydisorder1, Shender1982, orderbydisorder2, orderbydisorder3}, and fault-tolerant quantum computation~\cite{Chen2010}. Specifically, zero-energy degeneracy (ZED) plays a crucial role in topological phases~\cite{Kitaev_2001, Ivanov2001, chetan_nayak}, where it informs about the dimension of the ground state manifold -- thereby revealing the quantum information capacity of the topologically protected state.
Typically arising due to symmetry, zero-energy eigenstates have been recently studied in the celebrated PXP model in the context of thermalization~\cite{Turner2018}. These PXP zero-energy states exhibit several interesting properties, including area law entanglement~\cite{Lin2019}, stability in experimental setups~\cite{Karle2021}, and being building blocks for quasiparticle excitations~\cite{Lin2019, Surace2021}. Furthermore, certain lattice gauge theory models are endowed with constructible zero-energy many-body quantum scars~\cite{Debasish2021, Biswas2022, Sau2024}.

This raises questions about whether zero-energy subspaces in other models \cite{WeakUniversality} might demonstrate analogous fascinating characteristics.
Of particular interest would be an exploration of models with complex internal symmetries that have not been studied before. In this Letter, we investigate a class of nonintegrable bond-staggered (alternating) spin chains that possesses a non-Abelian internal symmetry. These models have attracted both theoretical~\cite{hida, hal_tasaki, hida2} and experimental~\cite{compound, Manaka1997Formation} interest due to their connection to Haldane spin chains and resemblance to the Su-Schrieffer-Heeger model.
Here, we employ character theory to obtain nullspace degeneracies within each symmetry sector, which we find to be scaling exponentially with the system size. Through symmetry considerations, we are able to construct a non-trivial exact zero-energy basis, serving as a few-body scarred subspace. This fixed-separation basis possesses low entanglement and is remarkably stable to experimental perturbations. These results push our current understanding of the dynamical stability of special states in non-Abelian chaotic models.

\textit{Model.}---
Consider a class of bond-staggered Heisenberg spin chains defined on a periodic lattice of $L$ sites,%
\begin{subequations}%
\begin{align}
  H & = \sum_{i=1}^L (- 1)^i \, \mathbf{S}_i \cdot \mathbf{S}_{i + 1} + V \label{eq:Ham}\\
  & = \frac{1}{2}\sum_i (- 1)^i \, \mathrm{SWAP}_{(i,i+1)} + V, \label{eq:HamSWAP}
\end{align}%
\end{subequations}%
where $\mathbf{S}_i = (S^x_i, S^y_i, S^z_i)$ is the vector of spin-half operators, and $V$ is any perturbation that preserves the symmetries of the model, as defined below. The presence of the perturbation term makes our results generic while allowing for investigations of stability against external disruptions. The second equality allows us to interpret the Hamiltonian as a collection of SWAP gates, which will prove useful in subsequent analysis. The model is endowed with a non-Abelian SU(2) total spin symmetry and the total spin conservation in any direction [U(1) symmetry]; without a loss of generality, we consider conservation of $S^z_\text{tot} = \sum_i S^z_i$. This forms the internal symmetry group $\mathcal G$. The Hamiltonian~(\ref{eq:Ham}) anticommutes with the translation operator $\tau$ and commutes with translation by two sites, $\tau^2$, and with bond-reflection symmetry, $\sigma$. Together, these form the spatial symmetry group of the model, the dihedral group $D_{L}$, illustrated in Fig.~\ref{fig:spatial_symmetries}(a). We term the $D_{2 L}$ group generated by $\tau$ and $\sigma$ the anti-symmetry group $\mathcal{A}$.

Previous research~\cite{Grabowski1995, Mukherjee2023} has implied that staggering the interaction term is responsible for breaking the integrability of the Heisenberg chain. We show this explicitly through level statistics for the periodic model with $V=0$ by resolving all its symmetries. We write the Hamiltonian in the fixed momentum magnon basis~\cite{Sandvik2010}, which resolves $S^z_\text{tot}$ and $\tau^2$. $\mathbf{S}^2$ is resolved by comparing the spectra of two consecutive values of $S^z_\text{tot}$, while $\sigma$ is diagonalized directly. We then plot the distribution of the ratio of consecutive level spacings $r$~\cite{Oganesyan2007} in Fig.~\ref{fig:spatial_symmetries}(b), which we find to match the prediction for the Gaussian orthogonal ensemble (GOE) of random matrix theory~\cite{Atas2013}. This implies the model is chaotic, strongly suggesting its nonintegrability. Moreover, the distribution exhibits level repulsion at $r=0$, while if the model had some remaining symmetries or was integrable, repulsion would be absent and the histogram would resemble the Poisson distribution.

\begin{figure}[tb]
  \includegraphics[width=1\columnwidth]{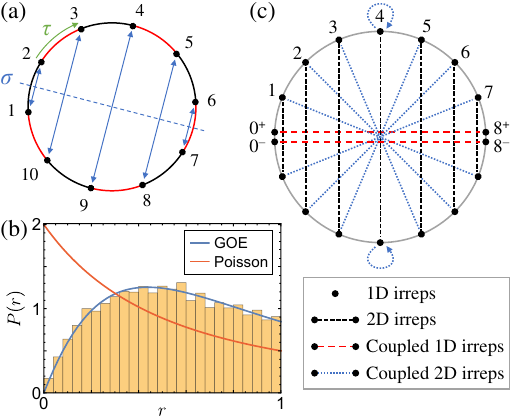}
  \caption{(a)~Spatial symmetries of the bond-staggered Heisenberg chain of size $L=10$: reflection symmetry $\sigma$ (blue), and translation $\tau$ (green). (b)~Level statistics for $L=22, S = 0, S^z_\text{tot} = 0$, showing a chaotic behavior of the ratio of consecutive level spacings $r$ (Gaussian orthogonal ensemble distribution). Integrable models follow the Poisson distribution. To improve the statistics, the results are gathered for all constant momenta sectors $0 < k < \pi$. (c)~Representation theory of $D_{2L}$ for $L=16$. Each representation labeled by integer momentum $\kappa$ is coupled by the anti-symmetry group $\mathcal{A}$ (red and blue lines). \label{fig:spatial_symmetries}}
\end{figure}

\textit{Zero-energy states enforced by the symmetry.}---
Through symmetry considerations, we calculate the lower bound on the nullspace dimension of the Hamiltonian and compare the result with exact diagonalization. The intuition for why the symmetry can inform about ZED is as follows. Consider any Hamiltonian $H$ that anticommutes with some unitary operator $A$, thus implying a symmetric spectrum ($E \mapsto -E$). Suppose that $A$ has only real eigenvalues, $\pm 1$, with multiplicities $m_{+}$ and $m_{-}$. In the eigenbasis of $A$, the Hamiltonian takes a block off-diagonal form, consisting of blocks of sizes $(m_{+} \times m_{-})$. Consequently, $H^2 = H^\dagger H$ becomes block-diagonal, separating into two sectors of $A$, with dimensions $(m_{+}\times m_{+})$ and $(m_{-}\times m_{-})$, both necessarily of the same rank. If there is an imbalance between $m_{+}$ and $m_{-}$, then the nullspace of $H^2$ (and consequently the nullspace of $H$) is at least $| m_{+} - m_{-} |$-dimensional~\cite{Sutherland1986}. Note that ZED may be larger if both sectors contain zero-energy states. However, if the model is chaotic, the lower bound is expected to be saturated.

This idea is refined when the anticommuting operator is part of a larger antisymmetry group.
We find that the $\mathcal{A}$-sectors are paired by the action of $H$ as (1)~one-dimensional irreducible representations labeled by momenta $k=0$ and $k=\pi$ with the same reflection eigenvalue $s=\pm 1$, and (2)~the two-dimensional irreps with momenta $k$ and $(\pi-k)$, with $k=2\pi \kappa/L$, where $\kappa = 1,\ldots,L/2-1$ [see Fig.~\ref{fig:spatial_symmetries}(c)].
Hence, ZED is bounded by
\begin{equation}
  \mathcal{Z} \ge \sum_{s=\pm 1} | m_{0,s} - m_{L/2,s} | + 2 {\sum_{\kappa=1}^{\lfloor{(L-2)/4}\rfloor}}| m_\kappa - m_{L/2 -\kappa} |
  \text{,}
  \label{eq:zed}
\end{equation}
where $m_{\kappa, s}$ are the multiplicities of the antisymmetry representation labeled by $(\kappa, s)$, with the $s$ label suppressed in the second sum as it is forced by $\kappa$. This is enhanced by internal symmetries, which cause the computation to be performed in each $\mathcal G$ sector separately.

We can calculate these symmetry-resolved multiplicities using character theory~\cite{Hamermesh2003, Sagan2001}.
The character function $\chi_\rho(g) = \tr(\rho(g))$ is a function that identifies a representation $\rho$.
Specifically, we take the character function for the representation on the full Hilbert space $\mathcal{H}$ and we use the Schur orthogonality relations for the natural inner product: the multiplicity of each irreducible representation within that full representation is the inner product of the irrep character and the character of the full representation.
Due to the product structure of the group we are interested in, we can write this full character as a function of $g\in \mathcal{A}$ and $\theta \in \mathcal{G}$.
This allows the inner product to be taken in two steps, first over the spatial part to produce $F_{\kappa,s}(\theta) = \langle \chi^\mathcal{A}_{\kappa,s}(g), \chi_{\mathcal{H}} (g, \theta) \rangle_{\mathcal{A}}$. Note that $\chi^\mathcal{A}_{\kappa,s}(\sigma^\nu \tau^\mu) = d_\kappa s^\nu \cos(2\pi\kappa \mu/L)$, where $d_\kappa \in \{1,2\}$ is the dimensionality of irrep $\kappa$~\footnote{See Appendix~A in the Supplemental Material for more details about character theory, tables of zero-energy degeneracies, and discussion of extra zero-energy states.}.
Second, over the internal symmetry group to extract the multiplicity of an individual representation,
\begin{align}
  m_{\kappa,s,S} = \left\langle \chi_S(\theta), F_{\kappa,s}(\theta)\right\rangle _{\mathcal{G}},
  \label{eq:multiplicity}
\end{align}
where $S$ labels the total spin sector. Note that one does not need to resolve $S^z_\text{tot}$, as it only acts inside each SU(2) representation.

Now, viewing $g$ as a permutation of sites, we can use its decomposition into cycles of sites,
\begin{equation}
  \chi_\mathcal{H}(g,\theta) = \mathrm{tr}_\mathcal{H} g\theta = \smashoperator{\prod_{\gamma \in \text{cycles of }g}} \mathrm{tr}_\gamma\!\left(\tau \theta\right)
  = \prod_l [\mathrm{tr}_{\mathcal{H}_l}(\tau\theta) ]^ {c_l(g)}, \label{eq:cycle_function}
\end{equation}
where $c_l(g)$ is the $l$-cycle index of $g$ which counts the number of cycles of length $l$ in $g$~\cite{Note1}.
The final trace appearing in Eq.~(\ref{eq:cycle_function}) is over a system of size $l$ (Hilbert space $\mathcal{H}_l$) and we call this the $l$-cycle character function.
Analogously to its use for the single-site translation, we use $\tau$ here to refer to the generator of the cyclic group within each cycle.

\begin{table}[t]
  {\footnotesize{\begin{tabular}{l|llllllllllll|l}
    \hline \hline
    %& $S \rightarrow$ &   &   &   &   &   &   &   &   &   &    & \\
    $L$ & $S = 0$ & 1 & 2 & 3 & 4 & 5 & 6 & 7 & 8 & 9 & 10 & 11 &
    % & 11& 12 & 13 & 14   &15&16&
    % &&&&&&&
    Total\\
    \hline
    % 2 & 1 & 1 &  &  &  &  &  &  &  &  &  &  &  &  &    &&& 4\\
    4 & 0 & 1 & 1 &  &  &  &  &  &  &  &  &  & 8\\
    6 & 3 & 3 & 1 & 1 &  &  &  &  &  &  &  &  & 24\\
    8 & 0 & \cellcolor{red!15}6 (2) & 2 & 1 & 1 &  &  &  &  &  &  &  & \cellcolor{red!15}44 (32)\\
    10 & 10 & 10 & 7 & 7 & 1 & 1 &  &  &  &  &  &  & 144\\
    12 & \cellcolor{red!15}2 (0) & 7 & 7 & 4 & 4 & 1 & 1 &  &  &  &  &  & \cellcolor{red!15}146 (144)\\
    14 & \cellcolor{red!15}37 (35) & 35 & 27 & 27 & 11 & 11 & 1 & 1 &  &  &  &  & \cellcolor{red!15}714
    (712)\\
    16 & \cellcolor{red!15}4 (0) & 14 & 14 & 14 & 14 & 6 & 6 & 1 & 1 &  &  &  & \cellcolor{red!15}516
    (512)\\
    18 & 126 & 126 & 102 & 102 & 52 & 52 & 15 & 15 & 1 & 1 &  &  &
    3224\\
    20 & 0 & 50 & 50 & 48 & 48 & 31 & 31 & 8 & 8 & 1 & 1 &  &
    2208\\
    22 & 462 & 462 & 390 & 390 & 225 & 225 & 85 & 85 & 19 & 19 & 1
    & 1 & 14136\\
    \hline \hline
  \end{tabular}}}
  \caption{Zero-energy degeneracy $\mathcal{Z}$ of the bond-staggered Heisenberg model as a function of total spin quantum number $S$ and system size $L$, from numerics. The numbers in brackets are from the character theory (absent if theory matches numerics). Extra zero-energy states are highlighted in red. \label{tab:degS}}
\end{table}

At this point, we specialize to the main internal symmetry of interest, SU(2). The character function for the irreducible representation of spin-$S$ is $\chi_{S}^\text{SU(2)}(\theta) = \sin[(2S+1) \theta] / \sin \theta$~\cite{Hamermesh2003}, and this forms a basis for the space of character functions. Pointwise operations on these produce a ring, known variously as the character ring or representation ring, and are equivalent to the fusion rules of SU(2)~\cite{Husemoller1966}. This equivalence takes the tensor product of representations to pointwise-multiplication of characters and direct sum to addition, producing equations such as $\chi^\text{SU(2)}_{1/2} \chi^\text{SU(2)}_{1/2} = \chi^\text{SU(2)}_0 + \chi^\text{SU(2)}_1$ from the fusion rules of SU(2).
The character function for the $l$-cycle evaluates to
\begin{equation}
  \mathrm{tr}_{\mathcal{H}_l}\!\left(\tau\theta\right) = 2 \cos l\theta = \chi^\text{SU(2)}_{l/2}(\theta) - \chi^\text{SU(2)}_{l/2-1}(\theta).
  \label{eq:char_fun}
\end{equation}
Finally, we perform polynomial-complexity arithmetic in this ring~\cite{Note1}, and combine Eqs.~(\ref{eq:zed}--\ref{eq:char_fun}) to compute the zero-energy degeneracies in any sector.

We compare the number of zero-energy eigenstates predicted by the character theory with the exact diagonalization results for $V=0$ in Table~\ref{tab:degS} and find an almost exact match, except for a few extra zero-energy states for small system sizes~\cite{Note1}.
This additional degeneracy is most likely accidental, occurring due to small symmetry sector size, rather than from an unresolved symmetry, for which we see no evidence.
The degeneracy can be lifted with the addition of a perturbation, such as $V = \lambda \sum_i (-1)^i \mathbf{S}_i \cdot \mathbf{S}_{i+3}$ for a generic perturbation strength $\lambda$, while respecting the symmetries.
Another interesting finding is a correspondence between multiplicities in sectors $\{S = L/2-2n, k, s\}$ and $\{S-1, \pi-k, -s\}$, which we observe in character theory results across all available system sizes (up to $L = 42$). We note however that this correspondence is a mathematical property coming from the arithmetic of the cycles in Eq.~(\ref{eq:char_fun}), and does not imply the existence of a simple operator that maps states between the two sectors.

\begin{figure}[tb]
  \includegraphics[width=1\columnwidth]{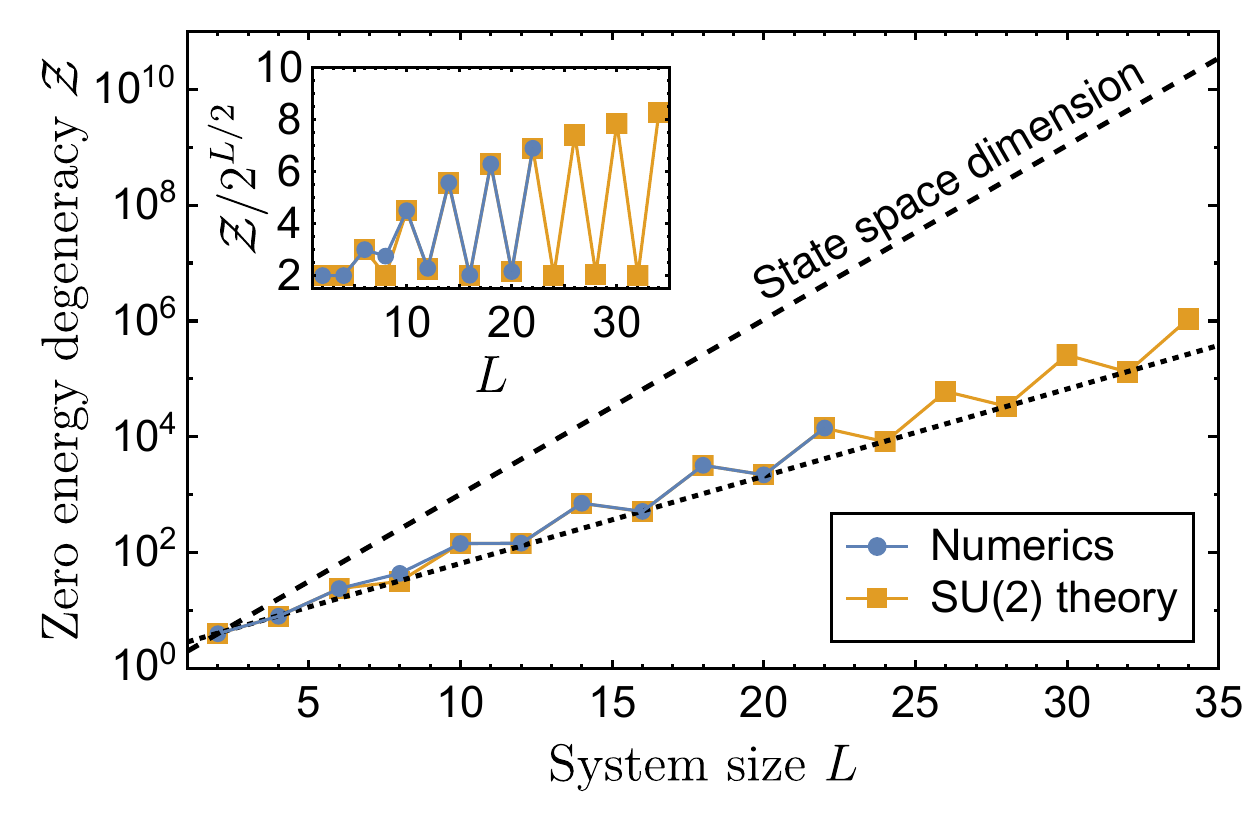}
  \caption{Zero-energy degeneracy $\mathcal{Z}$ as a function of the system size for the numerical results and character theory. The dashed line is the state space dimension of $2^L$, while the dotted line shows $2^{L/2+1}$. The inset shows $\mathcal{Z} / 2^{L / 2}$. \label{fig:zed}}
\end{figure}

We observe that the ZED grows exponentially with a rate half that of the state space dimension, $\ln \mathcal{Z} \sim \ln( 2^{L/2})$ (see Fig.~\ref{fig:zed}). Interestingly, a similar system size dependence has been seen previously in the PXP model~\cite{Turner2018}. The significant fraction of zero-energy states comes from the $k=0$ and $k=\pi$ sectors (the one-dimensional representations), which can be shown to follow $\mathcal{Z} \ge 2^{L/2+1}$~\footnote{See Appendix~B in the Supplemental Material for detailed estimation of the exponential behavior.}.
This simple prediction works especially well when $L$ is a power of two, for in that case the character-theoretic zeroes are only found in these sectors.
Similarly, as seen in the inset of Fig.~\ref{fig:zed}, for $L = 4n$ there are relatively few additional zeroes, while for $L = 4n+2$ other sectors contribute substantially more.

\textit{Exact zero-energy basis.}---
This exponentially large nullspace predicted by the character theory motivates our search for special analytically-tractable states.
Remarkably, we find that certain zero-energy states can be expressed in a closed form. In the following, we demonstrate how to construct a basis for two-magnon states, where a constant number of magnonic excitations is $N_\text{mag} = L/2 - S^z_\text{tot}$. The case of zero and one magnons is trivial, as there exists up to one zero-energy state per symmetry sector.

We use the computational basis with defined magnon positions, and assume the system size is divisible by 4. Using the action of $\mathcal{A}$ we partition this basis into orbits, $\mathrm{Orb}_\mathcal{A} \left(r_x\right) =  \{ g \ket{r_x} : \forall\, g \in \mathcal{A}\}$. Each orbit comes from a representative element $\ket{r_x}$ where the magnons have a defined separation $x \ge 1$, which uniquely identifies an orbit. Then within an orbit, we can produce a basis for the one-dimensional representations with character theory, creating symmetrized states
\begin{equation}
  \ket{r_x;\chi} = \frac{\sqrt{|\mathrm{Orb}_\mathcal{A}\!\left(r_x\right)\!|\,}}{|\mathcal{A}|}\sum_{g \in \mathcal{A}} \chi(g)\,g\!\ket{r_x}.
\end{equation}
Each orbit contains a so-called $0$ representation denoted $\chi_0$ which sets $\chi(\tau) = +1$ and gives an orbit-dependent sign for $\chi(\sigma)$. Every orbit, except for the orbit with separation $L/2$, has a so-called $\pi$ representation denoted $\chi_\pi$ which sets $\chi(\tau) = -1$ and again gives an orbit dependent sign for $\chi(\sigma)$. The reflection-parity signs are irrelevant for the fixed-separation states, justifying suppressing them from the notation.

Consider the action of $H$ in the form of Eq.~(\ref{eq:HamSWAP}) with $V=0$ upon some representative element $\ket{r_{2n}}$ with even separation, where $n=1,\ldots, L/4$. Applying a swap that is not incident on either magnon leaves the representative invariant, because there are just as many such positive and negative swaps. Hence, $\bra{r_{2n};\chi_0} H \ket{r_{2n}} = 0$. If we use a swap to move one of the magnons to the left we increase the separation, and for the other magnon, this happens if we swap it to the right. Because their initial separation is even, these swaps have opposite signs, hence $\bra{r_{2n+1};\chi_0} H \ket{r_{2n}} = 0$, since the contributions will destructively interfere.
A similar argument can be used to show that odd separation representatives satisfy $\bra{r_{2n+1};\chi_\pi} H \ket{r_{2n+1}} = 0$ and $\bra{r_{2n+2};\chi_\pi} H \ket{r_{2n+1}} = 0$. Combining these results, we obtain,
\begin{subequations}
\begin{align}
  H \ket{r_{2n};\chi_\pi} &= 0 & \text{for } n&=1,\ldots,L/4, \\
  H \ket{r_{2n+1}; \chi_0} &= 0 & \text{for } n&=1,\ldots,\lfloor{(L-2)/4}\rfloor
  \text{.}
\end{align}
\end{subequations}
Since the orbits are disjoint, this describes $(L/2 - 1)$ mutually orthogonal zero-energy eigenstates. The final zero-energy state comes from the uniform superposition in the computational basis, which is part of the maximum spin multiplet. In conclusion, when $L$ is a multiple of 4, we can construct an exact basis for all $L/2$ zero-energy eigenstates in the one-dimensional irreps of $\mathcal{A}$.

\textit{Entanglement and stability of the fixed-separation basis.}---
Given their analytical tractability, we inquire whether these zero-energy states are endowed with interesting physical properties, such as being interacting quantum scars. We discover that the states with fixed magnon separation exhibit an area-law von Neumann entanglement entropy scaling [see Fig.~\ref{fig:echo}(a)]~\footnote{See Appendix~C in the Supplemental Material for detailed calculations of the von Neumann energy of fixed-separation states.}, whose value approaches $\ln 2$ in the infinite volume limit. Conversely, if the separation scales linearly with the system size, the entropy behaves as expected for a generic state in a two-magnon sector, i.e., $\sim \ln L$.

\begin{figure}[tb]
  \includegraphics[width=1\columnwidth]{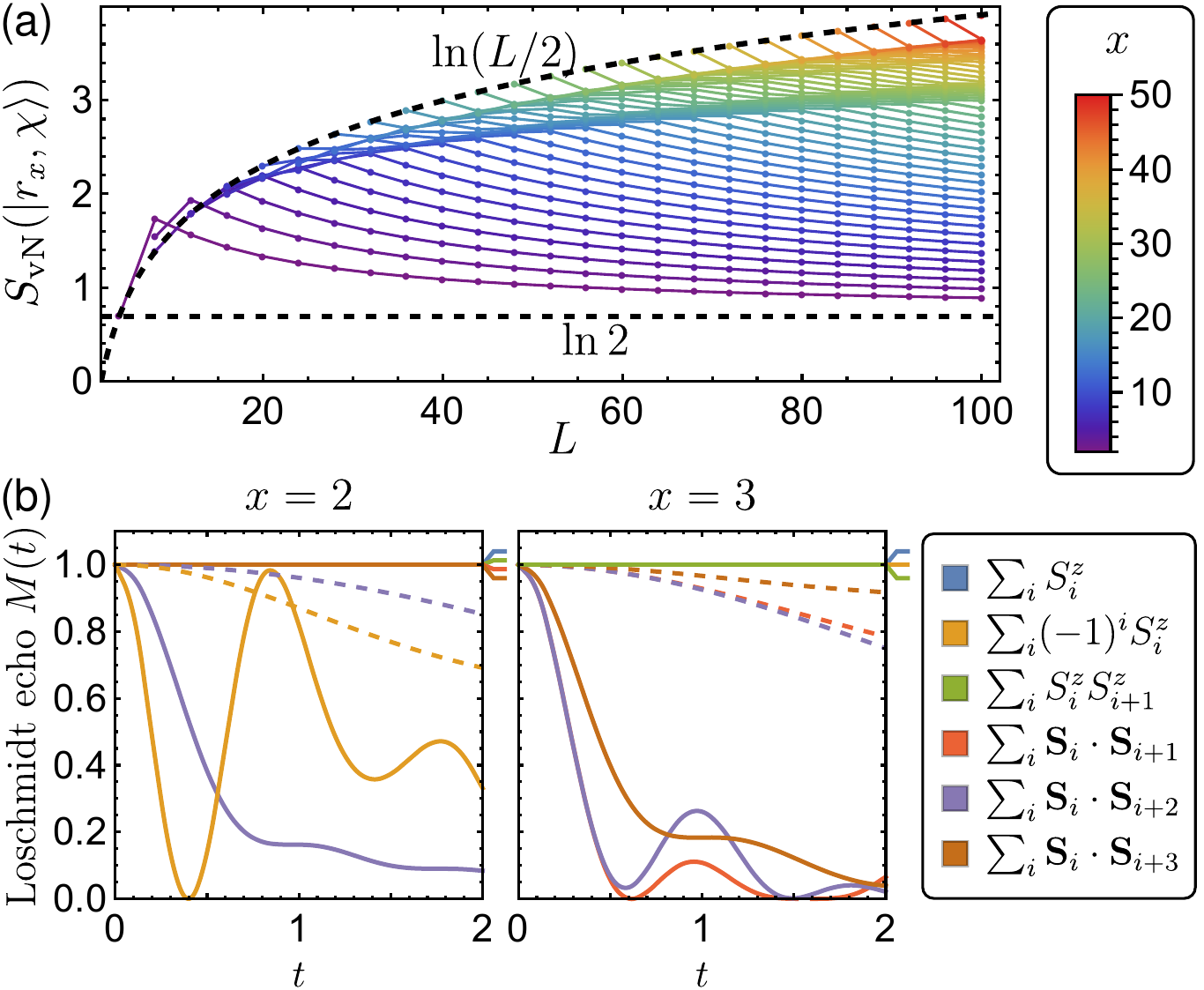}
  \caption{(a)~Entanglement entropy $S_\text{vN}$ for the states of constant separation $|r_x;\chi\rangle$. (b)~Loschmidt echo $M(t)$ for chain of size $L=32$, with magnon separations $x=2$ (left) and $x=3$ (right). The strength of perturbation terms is $\lambda=0.1$ (dashed lines) and $\lambda=1.0$ (solid lines).
  \label{fig:echo}}
\end{figure}

Surprisingly, the proposed basis is exceptionally stable to external perturbations. Consider the following terms that may naturally arise in experimental setups~\cite{Duan2002, Schafer2020, Lim2010, Weimer2010, Bernien2017}: external field, staggered field, Ising interaction, and extended Heisenberg interaction. We assess the stability of each state by measuring the Loschmidt echo at time $t$,
\begin{equation}
    M(t) = |\langle\psi| e^{i(H_0+\lambda V)t} e^{-i(H_0-\lambda V)t} |\psi\rangle|^2,
\end{equation}
where $V$ is the perturbation in question and $\lambda$ is its strength. We use representative examples of even/odd separation states to be those of $x=2$ and $x=3$, respectively. In each class the qualitative behavior is the same. As shown in Fig.~\ref{fig:echo}(b), we find that the basis is remarkably stable: states of even separations are only unstable to the staggered field and $\sum_i \mathbf{S}_i \cdot \mathbf{S}_{i+2}$, while odd-separation states are unstable to Heisenberg-type interaction terms. This implies that one can form a stable fixed-separation basis of zero-energy states, which can be used to encode some initial quantum information. This information is then preserved for a long period of time if the strength of the destabilizing perturbations is small. Interestingly, the Loschmidt echo for the staggered field perturbation shows an unusually strong revival at $t \approx 1, \lambda = 1.0$, implying that one can recover the initial state even if the additional term generally destroys the coherence of the state.

\textit{Discussion.}---
In this Letter, we have investigated a class of bond-staggered Heisenberg models. Despite their relative simplicity as two-body Hamiltonians, they exhibit rich physics, including a large number of zero-energy states. Through symmetry considerations, we provided the lower bound on the zero-energy degeneracy, showing that the nullspace grows exponentially with the system size. Furthermore, we constructed exact analytical formulas for up to two-magnon solutions using character theory. This fixed-separation zero-energy basis exhibits low entanglement and remarkable stability when subjected to perturbations that naturally occur in experiments.

This research opens up several exciting future directions. First, can a zero-energy basis composed solely of area-law states be constructed? States formed using Bethe ansatz methods usually exhibit low entanglement and are plausible candidates~\cite{Bibikov2018, Zhang2022}. Second, do exact constructions beyond two-magnon states exist? So far, we see no immediate way to extend the character-theory-based fixed-separation method to a larger number of magnons. Most interesting would be the existence of non-trivial closed-form states in the largest symmetry sector, or with volume law entanglement. Indeed, the presence of exponentially large nullspace may signify a hidden integrable structure, such as towers of states, possibly amendable to complementary analytical treatments. Finally, novel construction methods could be leveraged to build and stabilize zero-energy states, such as adaptive circuits~\cite{Ravindranath2023, Friedman2023}, where repeated measurements may be utilized to project perturbed states back into the nullspace manifold, mimicking the topological protection in surface codes, and making them valuable for fault-tolerant quantum computing.

\begin{acknowledgments}
\textit{Acknowledgments.}---
A.~P.,  B.~M., and M.~S.\@ were funded by the European Research Council (ERC) under the European Union's Horizon 2020 research and innovation programme (Grant Agreement No.\@ 853368). C.~J.~T.\@ is supported by an EPSRC fellowship (Grant Ref. EP/W005743/1). The authors acknowledge the use of the UCL High Performance Computing Facilities (Myriad and Kathleen), and associated support services, in the completion of this work. R.~M.\@ and H.~J.~C.\@ acknowledge funding from National Science Foundation Grant DMR 2046570 and the National High Magnetic Field Laboratory. The National High Magnetic Field Laboratory is supported by the National Science Foundation through NSF/DMR-1644779 and DMR-2128556 and the state of Florida.
\end{acknowledgments}

\bibliography{refs}

\clearpage

\title{Supplemental Material for ``Stable infinite-temperature eigenstates in SU(2)-symmetric nonintegrable models''}

\maketitle

\renewcommand\thefigure{S.\arabic{figure}}
\setcounter{figure}{0}
\renewcommand{\theequation}{S.\arabic{equation}}
\setcounter{equation}{0}
\renewcommand{\thetable}{S.\Roman{table}}
\setcounter{table}{0}

\section{Appendix A: Zero-energy degeneracy and detailed properties of the extra zero-energy states}

In this Appendix, we clarify more details about the character theory derivation of the zero-energy degeneracy and provide extended tables.

\subsection{Character theory details}

Let us start by explicitly writing the representations for the dihedral group $D_{2L}$, which is the anti-symmetry group $\mathcal{A}$ of our system (see Fig.~\ref{fig:spatial_symmetries} in the main text). It contains one-dimensional irreps labeled by $\kappa = 0, L/2$ and $s = \pm 1$ with representation
\begin{equation}
  \rho_{\kappa,s}(\tau) = \omega_\kappa \qquad
  \rho_{\kappa,s}(\sigma) = s,
\end{equation}
where $\omega_k = \exp(2\pi i\kappa/L)$ and $\rho$ is the linear representation. On the other hand, the two-dimensional irreps have the other allowed values of $\kappa = 1, 2, \dots, L/2{-}1$ with a matrix representation,
\begin{equation}
  \rho_\kappa(\tau) = \left(\begin{array}{cc} \omega_\kappa & 0 \\ 0 & \omega_\kappa^\ast \end{array}\right) \qquad
  \rho_\kappa(\sigma) = \left(\begin{array}{cc} 0 & 1 \\ 1 & 0 \end{array}\right).
\end{equation}
The characters are defined as traces of representation matrices, and hence it is easy to see that for 1D irreps $\chi_{\kappa,s}(\sigma^\nu \tau^\mu) = s^\nu \omega_\kappa^\mu$, while for 2D irreps $\chi_{\kappa}(\sigma^\nu \tau^\mu) = 2 \cos(2\pi\kappa\mu/L)$. Here, $\nu\in\{0,1\}$ and $\mu\in\{0,...,L-1\}$ denote the powers of the reflection and translation in the group element $g=\sigma^\nu \tau^\mu$. We can write this succinctly as $\chi_{\kappa,s}(\sigma^\nu \tau^\mu) = d_\kappa s^\nu \cos(2\pi\kappa\mu/L)$, where $d_\kappa \in \{1,2\}$ is the dimensionality of irrep $\kappa$ and for 2D irreps we take $s=+1$.

Now, we turn to the discussion of characters for the internal symmetry of the model, SU(2).
Consider the character function for the system of size $l$ (Hilbert space $\mathcal{H}_l$) from Eq.~\eqref{eq:cycle_function} in the main text.
The trace can be expressed in some orthonormal product basis, for $\theta$ an element of the internal symmetry and $\tau$ generates a cycle of length $l$,
\begin{subequations}
\begin{align}
  \mathrm{tr}_{\mathcal{H}_l}\!\left(\tau\theta\right)
  &=\smashoperator{\sum_{\alpha_1,\dots,\alpha_l}} \bra{\alpha_1}\cdots\bra{\alpha_l} \tau \theta \ket{\alpha_1}\cdots\ket{\alpha_l}\\
  &= \smashoperator{\sum_{\alpha_1,\dots,\alpha_l}} \bra{\alpha_1} \rho_{\frac{1}{2}}(\theta) \ket{\alpha_2} \cdots \bra{\alpha_l} \rho_{\frac{1}{2}}(\theta) \ket{\alpha_1}\\
  &= \sum_{\alpha_1} \bra{\alpha_1} \rho_{\frac{1}{2}}(\theta)^l \ket{\alpha_1} = \mathrm{tr} [ \rho_{\frac{1}{2}}(\theta)^l ]\text{,}
\end{align}%
\label{eq:trrho}%
\end{subequations}%
where $\rho_{\frac{1}{2}}(\theta)$ is the spin-$\frac{1}{2}$ representation of SU(2) for a single site, and $\ket{\alpha_i}$ is the basis for site $i$.
The first step is to use $\tau$ to permute the product basis labels, then $\theta$ is split as a product operator acting as $\rho_{\frac{1}{2}}(\theta)$ on each site.
The formula is then distributed into a product across sites, which is equivalent to taking the trace of a matrix power.

Now we can note that any spin-1/2 SU(2) matrix can be transformed into a diagonal form $\rho_\frac{1}{2}(\theta) = \diag(e^{i\theta},e^{-i\theta})$, giving us a straightforward way to compute the trace and characters,
\begin{equation}
    \mathrm{tr}_{\mathcal{H}_l}\!\left(\tau\theta\right) = 2 \cos l\theta = \chi_{l/2}(\theta) - \chi_{l/2-1}(\theta),
\end{equation}
where the characters for irreducible representations of spin-($l/2$) are $\chi_{l/2}^\text{SU(2)} = \sin[(l + 1) \theta] / \sin \theta$.

\begin{figure}[tb]
  \includegraphics[width=1\columnwidth]{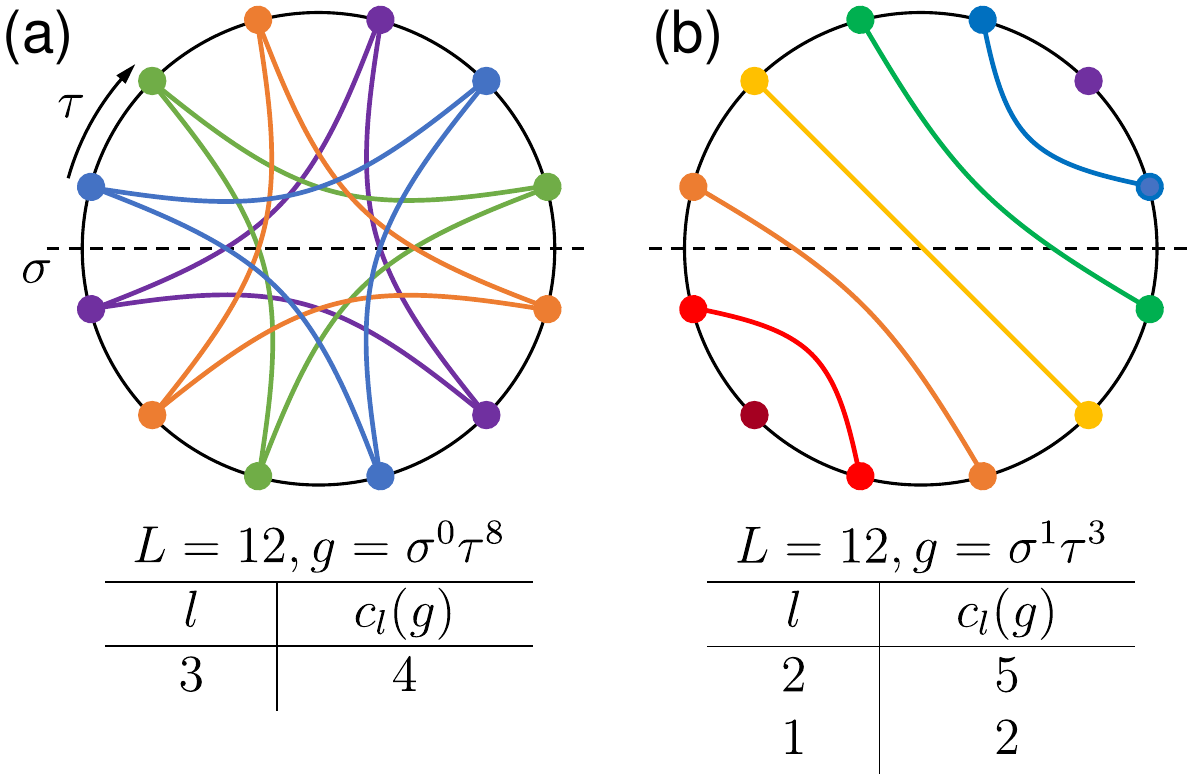}
  \caption{Diagram showing examples of $l$-cycles for a system of size $L=12$, for (a)~$g=\sigma^0 \tau^8$, and (b)~$g=\sigma^1 \tau^3$. Each color corresponds to one $l$-cycle. The corresponding cycle lengths $l$ and their counts $c_l(g)$ are listed in the tables below the panels.
  \label{fig:cycles}}
\end{figure}

Now using all these details, we explicitly write the expression for the multiplicities of the sector $\{\kappa,s,S\}$ [Eq.~\eqref{eq:multiplicity} in the main text]:
\begin{align}
   m_{\kappa,s,S} &= \langle \chi_S(\theta), \langle \chi_{\kappa,s}(g), \chi_\mathcal{H}(g,\theta) \rangle_\mathcal{A} \rangle_\mathcal{G}\\
   &=\int_0^{2\pi} \frac{\sin^2 \theta\, d\theta}{\pi} \chi_S^\ast (\theta) \sum_{g\in \mathcal A} \frac{1}{2L} \chi_{\kappa,s}^\ast (g) \chi_\mathcal{H} (g,\theta),\nonumber
\end{align}
where the integral element is on the torus of SU(2). Finally,
\begin{align}
    m_{\kappa,s,S} &= \int_0^{2\pi} \frac{\sin^2 \theta\, d\theta}{2 L \pi} \frac{\sin(2S+1)\theta}{\sin\theta} \label{eq:ZEDfull}\\
    &\quad \times \sum_{\mu = 0}^{L-1} \sum_{\nu = 0}^1 d_{\kappa} s^\nu \cos\frac{2\pi \kappa \mu}{L} \prod_l (2 \cos \theta l)^{c_l(g)}.\nonumber
\end{align}
This can be evaluated explicitly after first obtaining $l$-cycles and their counts $c_l(g)$, as follows. Two representative examples of how to calculate $\{l, c_l(g)\}$ are shown in Fig.~\ref{fig:cycles}, with panel (a) presenting a translation-only case ($\nu=0$), and panel (b) showing a case with translation and reflection ($\nu=1$). For $\nu=0$, there are $c_l(g)=\text{gcd}(L,\mu)$ cycles of length $l=L/\text{gcd}(L,\mu)$. For $\nu=1$, there are $c_2(g)=L/2-1$ cycles of length $2$ and $c_1(g)=2$ cycles of length $1$.

\subsection{Zero-energy degeneracies}

We show the detailed tables for the number of zero-energy states. Specifically, Table~\ref{tab:appZED} shows the comparison between numerical results and character theory up to $L=34$, as a function of total spin quantum number $S$ for a fixed value of $S^z_\text{tot}$, similarly to Table~\ref{tab:degS} in the main text.

\begin{table*}[tb]
  {\footnotesize{\begin{tabular}{l|lllllllllllllllll|l}
    \hline \hline
    & $S \rightarrow$ &  &  &  &  &  &  &  &  &  &  &  &  &  &    &&& \\
    $L \downarrow$ & 0 & 1 & 2 & 3 & 4 & 5 & 6 & 7 & 8 & 9 & 10 
    & 11& 12 & 13 & 14   &15&16&
    % &&&&&&&
    Total\\
    \hline
    2 & 1 & 1 &  &  &  &  &  &  &  &  &  &  &  &  &    &&& 4\\
    4 & 0 & 1 & 1 &  &  &  &  &  &  &  &  &  &  &  &    &&& 8\\
    6 & 3 & 3 & 1 & 1 &  &  &  &  &  &  &  &  &  &  &    &&& 24\\
    8 & 0 & \cellcolor{red!15}6 (2) & 2 & 1 & 1 &  &  &  &  &  &  &  &  &  &    &&& \cellcolor{red!15}44 (32)\\
    10 & 10 & 10 & 7 & 7 & 1 & 1 &  &  &  &  &  &  &  &  &    &&& 144\\
    12 & \cellcolor{red!15}2 (0) & 7 & 7 & 4 & 4 & 1 & 1 &  &  &  &  &  &  &  &    &&& \cellcolor{red!15}146 (144)\\
    14 & \cellcolor{red!15}37 (35) & 35 & 27 & 27 & 11 & 11 & 1 & 1 &  &  &  &  &  &  &    &&& \cellcolor{red!15}714
    (712)\\
    16 & \cellcolor{red!15}4 (0) & 14 & 14 & 14 & 14 & 6 & 6 & 1 & 1 &  &  &  &  &  &    &&& \cellcolor{red!15}516
    (512)\\
    18 & 126 & 126 & 102 & 102 & 52 & 52 & 15 & 15 & 1 & 1 &  &  &  &  &    &&&
    3224\\
    20 & 0 & 50 & 50 & 48 & 48 & 31 & 31 & 8 & 8 & 1 & 1 &  &  &  &    &&&
    2208\\
    22 & 462 & 462 & 390 & 390 & 225 & 225 & 85 & 85 & 19 & 19 & 1
    & 1 &  &  &    &&& 14136\\
    24 & (0) & (132) & (132) & (167) & 167 & 110 & 110 & 44 & 44 & 10 &
    10 & 1 & 1 &  &    &&& (8224)\\
    26 & (1716) & (1716) & (1485) & (1485) & (935) & (935) & 418 & 418 &
    126 & 126 & 23 & 23 & 1 & 1 &    &&& (60688)\\
    28 & (0) & (459) & (459) & (572) & (572) & (447) & (447) & (208) & 208 &
    71 & 71 & 12 & 12 & 1 & 1   &&& (33680)\\
    30 & (6435) & (6435) & (5671) & (5671) & (3803) & (3803) & (1911) & (1911)
    & (697) & 697 & 175 & 175 & 27 & 27 & 1   &1&& (257616)\\
    32 & (0) & (1430) & (1430) & (2002) & (2002) & (1638) & (1638) & (910) &
    (910) & (350) & 350 & 90 & 90 & 14 & 14   &1&1& (131072)\\
    34 & (24310) & (24310) & (21736) & (21736) & (15288) & (15288) & (8372) &
    (8372) & (3500) & (3500) & (1080) & (1080) & 232 & 232 & 31 & 31 & 1 & (1085760) \\
    \hline \hline
  \end{tabular}}}
  \caption{Numerical results for the size of nullspace of the bond-staggered Heisenberg model as a function of total spin quantum number $S$ and system size $L$. The numbers in brackets are from the SU(2) theory. When the number in the bracket is absent, the theory bound matches the numerics. Extra zero-energy states are highlighted in red. \label{tab:appZED}
  }
\end{table*}

Next, we list the symmetry properties of the extra zero-energy states -- states that are not predicted by the character theory in Table~\ref{tab:appextrazeros}. Except for the special case of $L=8$, the other extra zero-energy states are all within the $S=0$ sector; there exist an equal number of extra zero states with momentum $k=0$ and with momentum $k=\pi$, while their reflection $s$ is constant for a fixed system size. These extra zeros vanish when an additional term of $\lambda \sum_i(-1)^i \mathbf{S}_i \cdot \mathbf{S}_{i+3}$ is added to the Hamiltonian with a generic nonzero coefficient $\lambda$. This term does not change the symmetry properties of the Hamiltonian and instead makes the system ``more chaotic'', i.e., its properties more closely resemble those of a random matrix.

\begin{table}[t]
    \setlength\tabcolsep{6pt}
    \begin{tabular}{lllll}
        \hline \hline
        $L$ & $S$ & $k$ & $\sigma$ & $\mathcal{Z}$\\
        \hline
        8 & 1 & $\pi / 4$ &  & 2\\
        &  & $3 \pi / 4$ &  & 2\\
        \hline
        12 & 0 & $0$ & $+ 1$ & 1\\
        &  & $\pi$ & $+ 1$ & 1\\
        \hline
        14 & 0 & $0$ & $+ 1$ & 1\\
        &  & $\pi$ & $+ 1$ & 1\\
        \hline
        16 & 0 & $0$ & $- 1$ & 2\\
        &  & $\pi$ & $- 1$ & 2\\
        \hline \hline
    \end{tabular}
    \caption{Symmetry properties of extra zero-energy eigenstates.
    $k$ is the momentum associated with translation by one. $\mathcal{Z}$ is the number of extra zero-energy states.\label{tab:appextrazeros}}
\end{table}

\section{Appendix B: Closed-form degeneracy bound from 1D representations}

Here, we show that one can derive that the zero-energy degeneracy must grow at least as fast as square root of the Hilbert space dimension, more precisely, $\mathcal{Z} \ge 2^{L/2+1}$.
We do this by producing a lower bound $\mathcal{Z}^\text{lb}_L$ for the number of zeros required by character theory when (1)~only considering the one-dimensional spatial representations and (2)~erasing all internal symmetries.
This simplifies the calculations enough to give a closed-form expression for the bound.
Note that including the two-dimensional representations and the effect of internal symmetries can only cause the zero-energy degeneracy to increase.
In practice, we observe in the full character theory computations [Eq.~\eqref{eq:ZEDfull}] that $\ln \mathcal{Z} \sim L / 2 \ln 2$, i.e., that the true number of zeroes differs by some subexponential factor.

Let us first calculate the total character function when the internal symmetry is erased,
\begin{equation}
  \chi_\mathcal{H}(g) = \tr_\mathcal{H} g = {\prod_{\gamma \in \text{cycles of }g}} \tr_\gamma (\tau) = 2^{c(g)},
\end{equation}
where $\tau$ is the cyclic generator and $c(g)$ is the number of cycles of permutation $g$. This should be compared directly with Eq.~\eqref{eq:cycle_function} in the main text and Eq.~\eqref{eq:trrho}. Erasing the symmetries is equivalent to replacing $\rho_\frac{1}{2}(\theta)$ with an identity matrix of dimension $(2\times 2)$, yielding its trace as 2.

We will neglect all momentum sectors, other than $\kappa=0$ and $\kappa=L/2$ (one-dimensional representations), to obtain the lower bound of the zero-energy degeneracy $\mathcal{Z}_L^\text{lb}$,
\begin{equation}
  \mathcal{Z}_L \ge \mathcal{Z}_L^\text{lb} := \sum_{s = \pm 1} \left|\langle \chi_{0,s}, \chi_{\mathcal{H}_L} \rangle - \langle \chi_{L/2,s}, \chi_{\mathcal{H}_L} \rangle \right|.
\end{equation}
We can split the group into two cosets: a cyclic group coset with group elements $\tau^\mu$ and a reflection coset with group elements $\sigma\tau^\mu$. Likewise, the character inner products
\begin{equation}
    \langle \chi_{\kappa,s}, \chi_{\mathcal{H}_L} \rangle = \frac{1}{2L}\sum_{g\in\mathcal{A}} \chi_{\kappa,s}^\ast (g) 2^{c(g)}
\end{equation}
are split into two terms, one for each coset,
\begin{align}
T_{(\kappa,s)} &:= \frac{1}{2L} \sum_{\mu=0}^{L-1} \chi_{\kappa,s}^\ast(\tau^\mu) 2^{c(\tau^\mu)}, \\
R_{(\kappa,s)} &:= \frac{1}{2L} \sum_{\mu=0}^{L-1} \chi_{\kappa,s}^\ast(\sigma \tau^\mu) 2^{c(\sigma\tau^\mu)}.
\end{align}

First, we start with the $R_{(\kappa,s)}$ reflection contributions as they are both easier to compute and supply the majority of zero-energy states,
\begin{subequations}
\begin{align}
  R_{(0,s)} - R_{(L/2,s)} &= \frac{s}{2L}\sum_{\mu=0}^{L-1} (1 {-} (-1)^\mu) 2^{c(\sigma\tau^\mu)} \\
  &= \frac{s}{L}\sum_{\mu=0}^{L/2-1} 2^{c(\sigma\tau^{2\mu+1})} =  s\, 2^{L/2},
\end{align}
\end{subequations}
since $c(\sigma \tau^{2\mu+1}) = L/2 + 1$.
Since the $T_{(\kappa,s)}$ contributions are independent of $s$ and the $R_{(\kappa,s)}$ contributions are proportional to $s$, we can already conclude that,
\begin{subequations}
\begin{align}
  \mathcal{Z}_L^\text{lb} &= \smashoperator{\sum_{s=\pm 1}} \left|R_{(0,s)} - R_{(L/2,s)} + T_{(0,s)} - T_{(L/2,s)}\right| \\
  &\ge 2 \left|R_{(0,s)} - R_{(L/2,s)} \right| = 2^{L/2+1}.
\end{align}
\label{eq:zlb_inequality}
\end{subequations}
This matches the observed behavior from exact diagonalization in Fig.~\ref{fig:zed} of the main text.

We can also understand the behavior of the cyclic translation $T_{(\kappa,s)}$ terms,
\begin{equation}
  T_{(0,s)} - T_{(L/2,s)} = \frac{1}{2L}\sum_{\mu=0}^{L-1} \big(1 {-} (-1)^\mu\big)\, 2^{c(\tau^\mu)}.
  \label{eq:Tcontributions}
\end{equation}
If we ignore the signs under the sum, we would get a cycle index for the cyclic group on $L/2$ elements and if we take the $\mu$ even contributions these are permutations within the even and odd subsets -- hence these terms are equivalent to a cycle index for the cyclic group on $L/2$ elements.
So we can take a linear combination of these two to find,
\begin{equation}
  T_{(0,s)} - T_{(L/2,s)} = \frac{1}{L}\sum_{d\mid L} \varphi(d) 2^{\frac{L}{d}} - \frac{1}{L}\sum_{d \mid \frac{L}{2}} \varphi(d) 2^{\frac{L}{d}},
\end{equation}
where $\varphi(d)$ is the Euler's totient function (counting the integers smaller than $d$ that are relatively prime to $d$) and $d \mid L$ means that the sum is over positive $d$ which divide $L$.
Let $L=2^j Q$ where $Q$ is odd and let $d =2^j q$.
We can use these new quantities to rewrite the above and to bound it from above,
\begin{subequations}
\begin{align}
  T_{(0,s)} - T_{(L/2,s)} &= \frac{1}{L} \sum_{q|Q} \varphi(2^j q) \, 2^{Q/q} \\
  &\le 2^{Q}  \frac{1}{L} \sum_{q|Q} \varphi(2^j q)  \\
  &\le 2^{Q} \frac{1}{L} \sum_{d|L} \varphi(d) = 2^{Q} \le 2^{L/2},
\end{align}
\end{subequations}
using Gauss' divisor sum theorem. This is enough to make the previous inequality for $\mathcal{Z}_L^\text{lb}$ Eq.~\eqref{eq:zlb_inequality} into an equality, since the $T_{(\kappa,s)}$ terms are always smaller than the $R_{(\kappa,s)}$ terms.

To summarise, we have shown that $\mathcal{Z}_L \ge 2^{L/2+1}$, which agrees well with the numerics of Fig.~\ref{fig:zed} in the main text. Specifically, we get an exact match when $L$ is a power of two, as the zero-energy states in this case only exist in one-dimensional irreducible representations ($k=0$ and $k=\pi$ sectors). The match is also especially good when $L=4n$, where we find relatively few additional zeroes. For $L=4n+2$, two-dimensional irreps ($0<k<\pi$) contribute substantially more zeros, although the number is subexponential.

\section{Appendix C: Entanglement of the fixed-separation zero-energy basis}

In this section, we show how to analytically calculate the entanglement of the zero-energy states of the proposed fixed-separation basis. Von Neumann entanglement entropy is defined for a bipartition into region $A$ and its complement $B$ as $S_\text{vN} = - \tr(\rho_A \ln \rho_A)$, where $\rho_A$ is the reduced density matrix of $A$. We specialize to the case of half-chain entanglement entropy, where region $A$ is a half of the system.

Let us construct $\rho_A = \tr_B ( | \psi \rangle \langle \psi | )$, which has a block structure, each block associated with zero, one, or two magnons. We consider an unnormalized state with separation $x$, $|r_x; \chi \rangle$, as this makes our considerations easier, and represent the state in the computational basis. In the following, only separations $x < \lfloor L/4 \rfloor$ are considered. The zero-magnon sector is a $1\times 1$ matrix, with an element $(L/2-x)$. The two magnon sector is a $(L/2-x)\times (L/2-x)$ matrix, filled only with ones. This can be checked to give one nonzero eigenvalue of $(L/2-x)$. Finally, the one magnon sector is an identity matrix of size $(2x)\times(2x)$. To normalize the state we divide $\rho_A$ by its trace, which in this case is $\tr \rho_A = L$.

\begin{figure}[tb]
  \includegraphics[width=1\columnwidth]{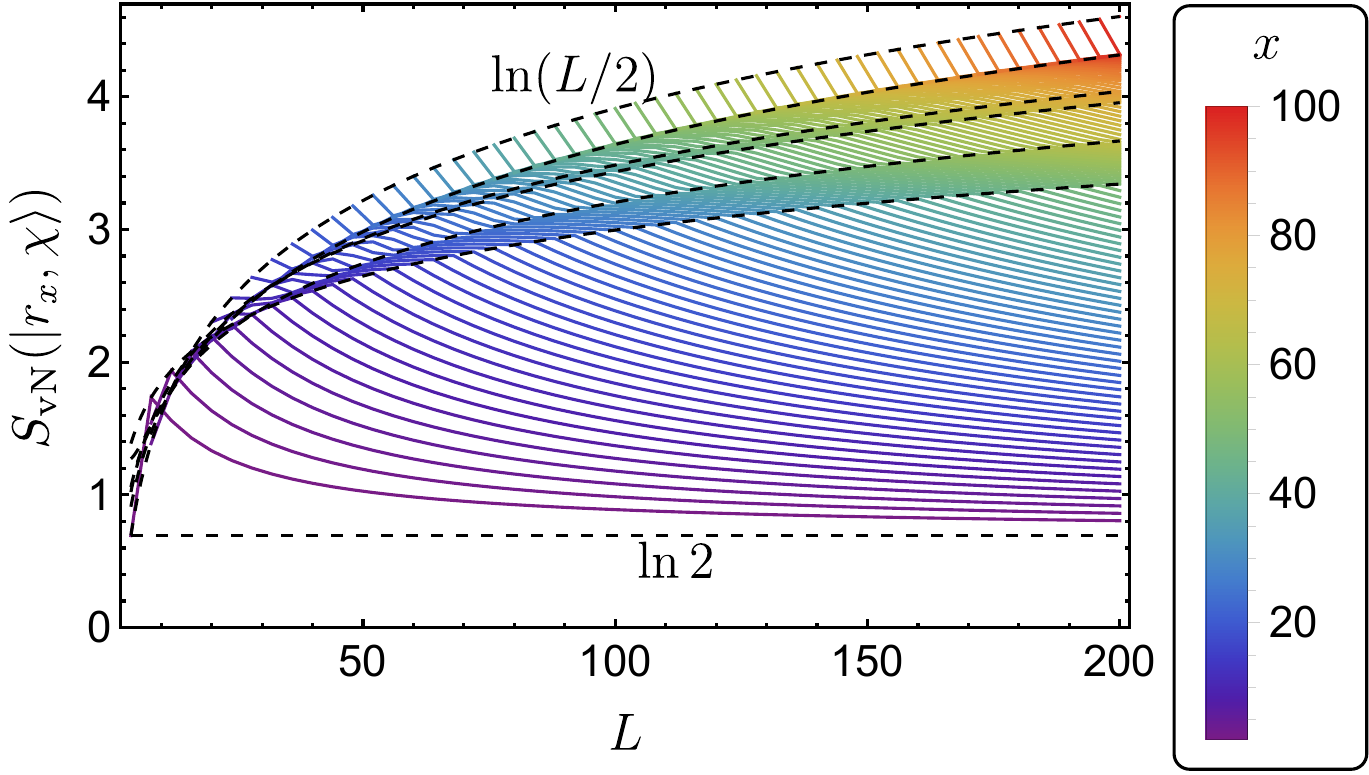}
  \caption{Entanglement entropy $S_\text{vN}$ for the states $|r_x;\chi\rangle$. Dashed black lines show the predictions from Eqs.~(\ref{eq:ent_pred1})--(\ref{eq:ent_pred2}).
  \label{fig:ent_detail}}
\end{figure}

Hence, the entanglement entropy of the $|r_x; \chi \rangle$ state is
\begin{equation}
    S_\text{vN}(x) = -2 \, \frac{L/2-x}{L} \ln \frac{L/2-x}{L} - 2 x \, \frac{1}{L} \ln \frac{1}{L}.
\end{equation}
Note that if we fix the separation $x$ and increase the system size, this expression tends to $\ln 2$ independently of the separation. On the other hand, if the separation is assumed to be proportional to the system size $x = \alpha L$, then the entropy diverges as $S_\text{vN} \sim 2 \alpha \ln L$. For comparison, a generic state in the two-magnon sector would have an entanglement proportional to $\ln L$.

\begin{figure}[t]
  \includegraphics[width=1\columnwidth]{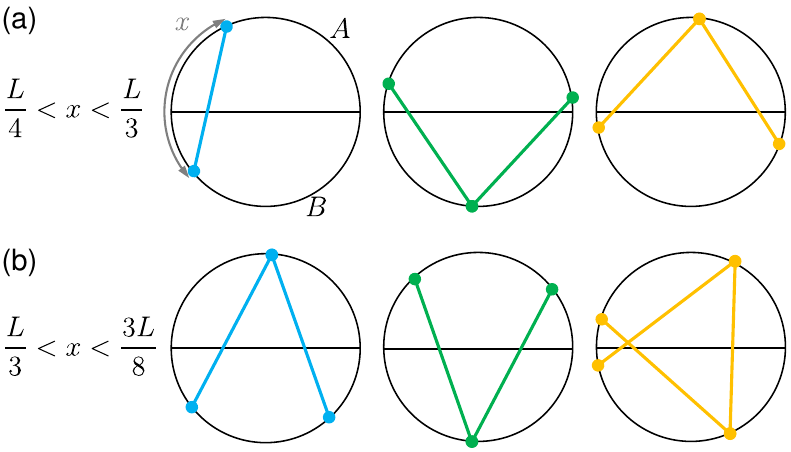}
  \caption{Structure of the one-magnon sector of the reduced density matrix for separations of (a)~$L/4 < x < L/3$ and (b)~$L/3 < x < 3 L/8$. The chain is represented as the circle, which is separated into subsystems $A$ and $B$. A fixed-separation state $\ket{r_x}$ is represented as two dots (magnon positions) connected by a line. Each diagram is then associated with matrix elements of the one-magnon sector of the reduced density matrix.
  \label{fig:rdm}}
\end{figure}

For $x > L/4$ the structure of $\rho_A$ is more complicated. The one magnon sector is now a sparse matrix of size $(L / 2) \times (L / 2)$, with non-zero diagonal and off-diagonal elements, which connect the $\rho_A$ basis states as shown in Fig.~\ref{fig:rdm}. Specifically, for each separation, one needs to consider three types of diagrams, each corresponding to elements of the reduced density matrix. For example, in Fig.~\ref{fig:rdm}(a) ($L/4 < x < L/3$), the first diagram gives a diagonal contribution for basis states when the magnon is distance $\in (2x-L/2, L/2-x)$ from the subsystem edge, while the second diagram gives a diagonal and off-diagonal contribution for basis states for which the magnon's distance to the edge of subsystem A is below $(2x-L/2)$.
The structure changes at special separations of $x = (k-1)L/(2k)$, with $k = 1, 2, ..., L/2-1$, when the diagrams in Fig.~\ref{fig:rdm} change. This leads the entropy curves to have non-analytic behavior in Fig.~\ref{fig:ent_detail}.

\begin{figure}[t]
  \includegraphics[width=1\columnwidth]{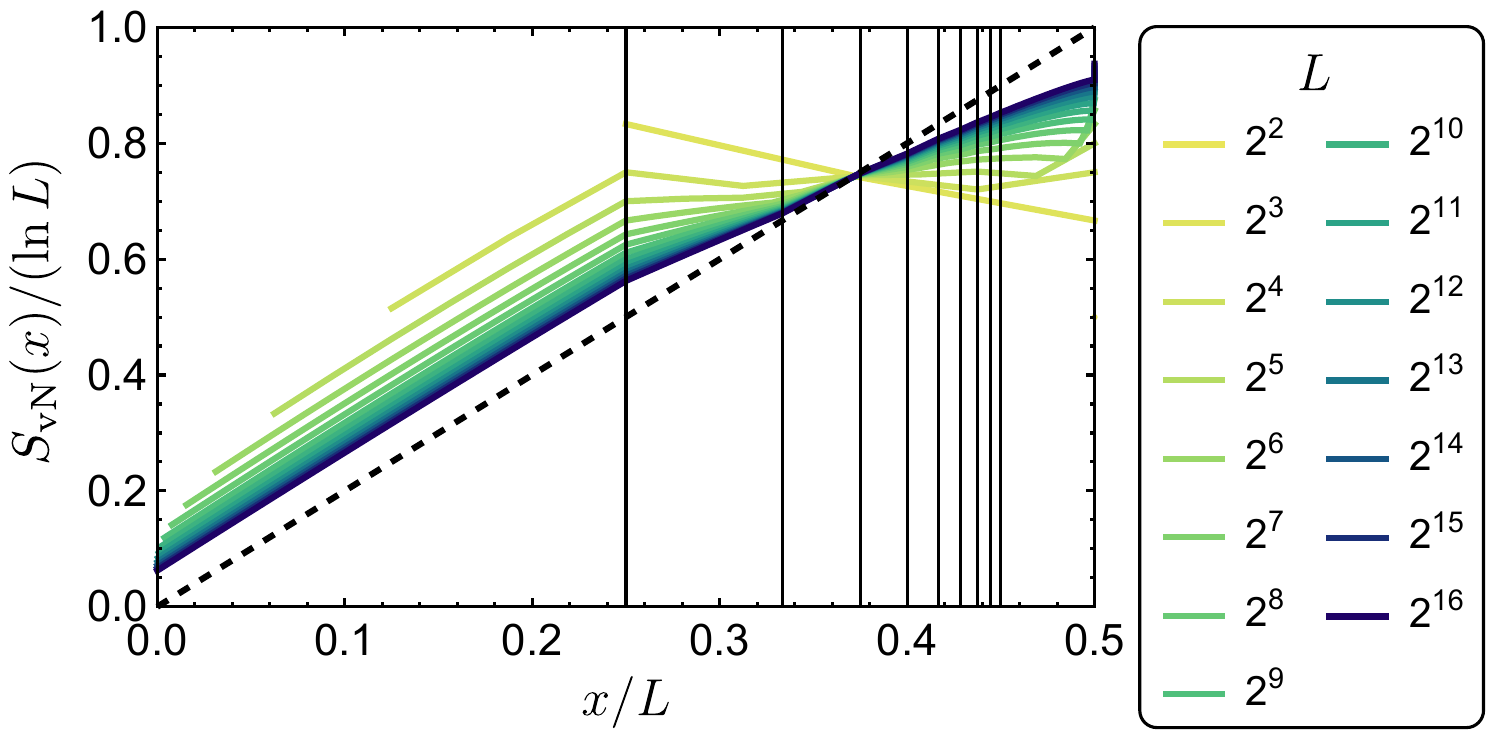}
  \caption{Rescaled entanglement entropy $S_\text{vN}$ for the states $|r_x;\chi\rangle$ as a function of the separation ratio, $x / L$. Special separations of $x = (n-1)L/(2n)$, with $n = 1, ..., 9$ are shown as black vertical lines. The infinite volume limit of $S_\text{vN}(x)/\ln L = 2x/L$ is designated by a dashed line.
  \label{fig:ent_detail_inf}}
\end{figure}

The entropy for some of these special separations can be derived to be
\begin{align}
    S_\text{vN}\biggl(x = \frac{L}{4}  \biggr) &= \frac{1}{2} \ln L - \ln 2, \label{eq:ent_pred1}\\
    S_\text{vN}\biggl(x = \frac{2L}{6} \biggr) &= \frac{2}{3} \ln L + \frac{1}{3} \ln \frac{3}{2},\\
    S_\text{vN}\biggl(x = \frac{3L}{8} \biggr) &= \frac{3}{4} \ln 2L - \frac{\sqrt{5}}{4} \coth^{-1} \frac{3}{\sqrt{5}},\\
    S_\text{vN}\biggl(x = \frac{4L}{10}\biggr) &= \frac{4}{5}\ln L-\frac{1}{5} \ln \frac{27}{10},\\
    S_\text{vN}\biggl(x = \frac{L}{2}-1\biggr) &\approx \ln L - 1 + \frac{8 \ln 2 - 2}{L},\\
    S_\text{vN}\biggl(x = \frac{L}{2}  \biggr) &= \ln \frac{L}{2}. \label{eq:ent_pred2}
\end{align}
These specific values match the kinks in the entanglement entropy plot in Fig.~\ref{fig:ent_detail}.
Interestingly, all the results approach the curve $S_\text{vN}(x) / \ln L = 2 x / L$ when $L$ is increased to infinity (see Fig.~\ref{fig:ent_detail_inf}), suggesting that in the thermodynamic limit, the entropy curve vs the separation is not only continuous but also smooth.

\clearpage

\end{document}